\documentclass[preprint,aps,amssymb,showpacs,superscriptaddress,nofootinbib]{revtex4}
\usepackage{graphicx}

\newcommand{\Case}[2]{{\textstyle \frac{#1}{#2}}}
\newcommand{\lP}{\ell_{\mathrm P}}

\begin{document}
%
\preprint{IMSc/2004/07/27}

\title{Effective Hamiltonian for Isotropic Loop Quantum Cosmology}

\author{Ghanashyam Date}
\email{shyam@imsc.res.in}
\affiliation{The Institute of Mathematical Sciences\\
CIT Campus, Chennai-600 113, INDIA.}

\author{Golam Mortuza Hossain}
\email{golam@imsc.res.in}
\affiliation{The Institute of Mathematical Sciences\\
CIT Campus, Chennai-600 113, INDIA.}

\begin{abstract}
For a class of solutions of the fundamental difference equation of
isotropic loop quantum cosmology, the difference equation can be
replaced by a differential equation valid for {\em all} values of the
triad variable. The differential equation admits a `unique' non-singular
continuation through vanishing triad. A WKB approximation for the
solutions leads to an effective continuum Hamiltonian. The effective
dynamics is also non-singular (no big bang singularity) and approximates 
the classical dynamics for large volumes. The effective evolution is
thus a more reliable model for further phenomenological implications of
the small volume effects.
\end{abstract}



\maketitle


\section{Introduction}

The singularities of classical general relativity, when specialized to
homogeneous, isotropic models, manifest as reaching zero physical
volume at finite synchronous time in the past. This in turn imply
unbounded growth of space-time curvature and of matter densities etc and
signals break down of the evolution equations at finite time in the past. 
It is widely believed that this feature of the classical theory will be
modified in a quantum theory of gravity and recent development of {\em
Loop Quantum Cosmology} (LQC)
\cite{cosmoI,cosmoII,cosmoIII,cosmoIV,IsoCosmo,LoopCosRev,Bohr}, 
corroborate this expectation \cite{InvScale,Sing}. 

The mechanism of `singularity avoidance' \cite{Sing} involves replacement 
of the classical evolution equation by a quantum one which is a {\em difference
equation} \cite{cosmoIV} thanks to the necessity of using holonomy operators
in the quantization of the Hamiltonian constraint in LQC. This equation
exhibits the property that the quantum wave function can be evolved
through zero volume unambiguously. In addition, the discreteness of the
triad operator (having zero eigenvalue) necessitates defining inverse 
triad operator (or inverse scale factor operator) \cite{InvScale} indirectly.
Thanks to the loop representation on the (non-separable) kinematical Hilbert 
space of LQC \cite{Bohr}, these operators get so defined as to have a
bounded spectrum implying only a bounded growth of curvatures/matter
densities. This is true for all allowed values of the ambiguity
parameters \cite{Ambig,ICGCAmbig}.

The non-separable structure of the Kinematical Hilbert space of LQC
however, also implies a huge set of solutions of the Hamiltonian
constraint (a continuous infinity in the gravitational sector alone).
Presumably, a suitable choice of physical inner product can be made to
cut down the size of the admissible solutions of the Hamiltonian
constraint. A choice of inner product however is not yet available. 
The exploration of Dirac observables is also at a preliminary stage 
\cite{HubbleOpr}. The general issue of whether or not the non-separable 
kinematical Hilbert space is mandatory, is currently an open issue 
\cite{Rovelli,Velhinho}. In the present work however we assume the current 
framework of LQC \cite{Bohr}.

Despite the open issues, it is possible to develop a WKB type semi-classical
approximation from which an {\em effective continuum Hamiltonian constraint}
can be deduced \cite{SemiClass,BIX}. This at once gives access to the usual
classical Hamiltonian methods to construct and analyze the quantum modified
space-time. This method relies on a {\em continuum approximation} 
\cite{FundamentalDisc} of the underlying difference equation to the 
Wheeler--DeWitt differential equation followed by the WKB ansatz for its
solutions. For large volume corresponding to classical regime, the continuum
approximation is always available, in fact as a requirement on quantization 
of the Hamiltonian operator. In this regime, the WKB ansatz naturally 
reproduces the classical Hamiltonian as the leading $o(\hbar^0)$ term. 
We would like to extend this method also to small volumes.

The large freedom offered by the non-separable structure of the kinematical 
Hilbert space can be exploited to propose a {\em restriction} to those 
solutions of the Hamiltonian constraint for which a continuum approximation
is valid for {\em all} volumes. One can then develop the effective classical
Hamiltonian constraint for all volumes and explore its consequences. 

In this work we develop such a picture and in comparison with the usual
FRW equations identify the effective density and pressure which includes
the contributions of quantum fluctuations of the geometry. Some elementary
consequences are also noted. Further implications for phenomenology are
discussed in separate papers \cite{GenericBounce,GenericInflation}. 

In section II, we detail the effective Hamiltonian constraint for
isotropic models. In section III, we discuss the qualitative features of
the corresponding dynamics namely, the possibility of `bounce' solutions
as well as solutions that could attempt to `pass through' the zero
volume and connect to the oppositely oriented isotropic universe. We
discuss what features of the quantum evolution are captured by the
effective classical evolution. In section IV, we summarize our conclusions 
and outlook.

\section{Effective Hamiltonian Constraint} 

The Kinematical Hilbert space \cite{Bohr} is conveniently described in 
terms of the eigenstates of the densitized triad operator,
\begin{equation}\label{TriadAction}
\hat{p} | \mu \rangle ~ = ~ \Case{1}{6} \gamma \lP^2 \mu | \mu
\rangle~~,~~ \langle \mu | \mu^{\prime} \rangle = \delta_{\mu \mu^{\prime}}
\end{equation}

The action of the volume operator on the triad basis states 
are given by
\begin{equation}
\hat{V} ~|\mu \rangle ~=~ {\left|\frac{1}{6}\gamma l_p^2 \mu \right|}^{\frac{3}{2}}
~|\mu \rangle  	~:=~ V_{\mu}~|\mu \rangle ~. 
\label{VolAction}
\end{equation}

In the isotropic context we have two classes to consider, namely
spatially flat and close models. The quantization of the
corresponding Hamiltonian operators is given in \cite{IsoCosmo}.
By introducing a parameter $\eta$ we can deal both classes together. The
values $\eta = 0$ and $\eta = 1$ will give the flat and the close
models respectively. 

The action of the gravitational Hamiltonian on the triad basis
states is then given by
\begin{eqnarray}
\hat{H}^{(\mu_0)}_{\text{grav}} |\mu \rangle & = & \left(\frac{3}{4
\kappa}\right)
{(\gamma^3 \mu_0^3 l_p^2)}^{-1} (V_{\mu+\mu_0} - V_{\mu-\mu_0})
\nonumber \\
& & \hspace{1.2cm} \left( ~e^{-i \mu_0\eta }|\mu + 4 \mu_0 \rangle ~-~ 
(2 + 4 \mu_0^2 \gamma^2\eta )|\mu \rangle ~+~ 
e^{i \mu_0\eta }|\mu - 4 \mu_0 \rangle ~\right)~ . 
\label{HGravAction}
\end{eqnarray}

Here, $\mu_0$ is a quantization ambiguity parameter which enters through
the fiducial length of the loops used in defining the holonomies. It is a real
number of the order of 1. Notice that the Hamiltonian connects states 
differing in their labels by $\pm 4 \mu_0$. This is a direct consequence of 
the use of holonomy operators which have to be used in the quantization of 
the Hamiltonian operator and is responsible for leading to a difference 
equation below. 

A general kinematical state $|s\rangle$, in the triad basis has the form 
\begin{equation}
|s \rangle ~=~ \sum_{\mu \in \mathbb{R}}~s_{\mu}~ |\mu \rangle ~~~~~~~~(\text{
sum over countable subsets})~.
\label{KinState}
\end{equation}

The Hamiltonian constraint of the classical theory is promoted as a
condition to define physical states, i.e.,
\begin{equation}
(\hat{H}^{(\mu_0)}_{\text{grav}} + \hat{H}^{(\mu_0)}_{\text{matter}}) |s \rangle ~=~ 0~.
\label{HamCons}
\end{equation}

In terms of $\tilde{s}_{\mu}:= e^{\frac{i \mu}{4}\eta} s_{\mu}$ and
The Hamiltonian constraint (\ref{HamCons}) translates into a difference
equation, 
\begin{eqnarray} \label{DifferenceEqn}
0 & = & A_{\mu + 4 \mu_0} \tilde{s}_{\mu + 4 \mu_0} ~-~ 
(2 + 4 \mu_0^2 \gamma^2 \eta) A_{\mu} \tilde{s}_{\mu} ~+~ 
A_{\mu - 4 \mu_0} \tilde{s}_{\mu - 4 \mu_0} \nonumber \\
& & \hspace{0.0cm} +~ 8 \kappa \gamma^2 {\mu_0}^3 
{\left(\frac{1}{6} \gamma l_p^2\right)}^{-\frac{1}{2}}
 H_{m}(\mu) \tilde{s}_{\mu} ~~~,~~~~~ \forall ~ \mu ~ \in \mathbb{R} \\
A_{\mu} & := & {|\mu + \mu_0|}^{\frac{3}{2}} -
{|\mu-\mu_0|}^{\frac{3}{2}}~~, \hspace{1.0cm} 
\hat{H}^{(\mu_0)}_{\text{matter}} |\mu\rangle ~ := ~ 
H_{m}(\mu) |\mu\rangle ~~ . \nonumber
\end{eqnarray}
$H_{m}(\mu)$ is a symbolic eigenvalue and we have assumed that
the matter couples to the gravity via the metric component and {\em not}
through the curvature component. In particular, $H_{m}(\mu = 0) =
0$. There are a few points about the above equation worth noting
explicitly.

Although $\mu$ takes all possible real values, the equation connects the 
$\tilde{s}_{\mu}$ coefficients only in steps of $4 \mu_0$ making it a 
difference equation for the coefficients. By putting $\mu := \nu + (4\mu_0) 
n,~ n \in \mathbb{Z}, ~\nu \in [0, 4 \mu_0)$, one can see that one has a 
continuous infinity of independent solutions of the difference equation, 
labelled by $\nu, ~ S^{\nu}_n := \tilde{s}_{\nu + 4 \mu_0 n}$. For each $\nu$ an
infinity of coefficients, $S^{\nu}_n$,  are determined by 2 `initial conditions'
since the order of the difference equation in terms of these coefficients is
2. Coefficients belonging to different $\nu$ are mutually decoupled. Since
the coefficients $A_{\mu}$ and the symbolic eigenvalues $H_{m}(\mu)$, 
both vanish for $\mu = 0$, the coefficient $\tilde{s}_0$ {\em decouples} 
from all other coefficients.

For large values of $\mu \gg 4 \mu_0$ ($n \gg 1$), which correspond to
large volume, the coefficients $A_{\mu}$ become almost constant (up to a
common factor of $\sqrt{n}$) and the matter contribution is also expected
similarly to be almost constant. One then expects the coefficients to 
vary slowly as $n$ is varied. This suggests interpolating these slowly varying
{\em sequences} of coefficients by slowly varying {\em functions} of the 
continuous variable $p(n) := \Case{1}{6} \gamma \lP^2 n$ \cite{SemiClass}. 
The difference equation satisfied by the coefficients then implies a {\em 
differential} equation for the interpolating functions which turns out to be 
{\em independent} of $\gamma$ and matches with the usual Wheeler--DeWitt
equation of quantum cosmology. This is referred to as a continuum approximation
\cite{FundamentalDisc}. This is of course what one expects if LQC 
dynamics is to exhibit a semi-classical behavior. While admissibility of
continuum approximation is well motivated for large volume, one also
expects it to be a poor approximation for smaller Planck scale volumes.

%

This logic is valid when applied to any {\em one} of the solutions 
${S^{\nu}_n}$. Thanks to the non-separable structure of the Hilbert
space, we have an infinity of solutions of the Hamiltonian constraint.
Although ${S^{\nu}_n}$ are uncorrelated for different $\nu$, nothing
prevents us from {\em choosing} them to be suitably correlated. In
effect this amounts to viewing $\tilde{s}_{\mu}$ themselves as {\em
functions} of the continuous variable $\mu$ and stipulating some
properties for them. In the absence of a physical inner product, we
don't have any criteria to select the class of solution. It is then
useful to study properties of classes of solutions of the Hamiltonian
constraint. 

The class that we will concentrate on is the class of {\em slowly
varying} functions. For these we will be able to have a continuum
approximation leading to a differential equation. Making a WKB
approximation for this differential equation, we will read-off the
effective classical Hamiltonian constraint. In anticipation of making
contact with a classical description, we will use the dimensionful
variable $p(\mu) := \Case{1}{6} \gamma \lP^2 \mu $ as the continuous
variable. Correspondingly, we define $p_0 := \Case{1}{6} \gamma
\lP^2 \mu_0$ which provides a convenient {\em scale} to demarcate
different regimes in $p$. We also use the notation: $\psi(p(\mu)) :=
\tilde{s}_{\mu}$. Now the definition of a slowly varying function is
simple: $\psi(p)$ is {\em locally slowly varying around $q$} if $\psi(q +
\delta q) \approx \psi(q) + \delta q \Case{d \psi}{dq} + \Case{1}{2}
\delta q^2 \Case{d^2 \psi}{dq^2} + \cdots$ with successive terms smaller
than the preceding terms, for $\delta q \lesssim 4 p_0 $. It is slowly varying
if it is locally slowly varying around every $q \in \mathbb{R}$
\cite{FundamentalDisc}. Note that even an exponentially rising function 
can be locally slowly varying if the exponent is sufficiently small.

To explore the possibility of slowly varying solutions of the difference
equation (\ref{DifferenceEqn}), consider the difference equation
more explicitly. For $\nu \in (0, 4\mu_0)$, putting $S_{n}(\nu) := 
\tilde{s}_{\nu + 4 \mu_0 n}, ~ A_n(\nu) := A_{\nu + 4 \mu_0 n}$ and 
momentarily ignoring the matter term for notational simplicity, the difference
equation (\ref{DifferenceEqn}) can be written as,
\begin{equation}
S_{n + 2}(\nu) = \left[\left(2 + 4 \mu_0^2 \gamma^2 \eta\right)
\frac{A_{n + 1}(\nu)}{A_{n + 2}(\nu)}\right] S_{n + 1}(\nu) + \left[\frac{-
A_n(\nu)}{A_{n + 2}(\nu)}\right] S_n(\nu) \label{DiffEqn}
\end{equation}
Its general solution can be written as $S_n(\nu) = S_0(\nu) \rho_n(\nu)
+ S_1(\nu) \sigma_n(\nu)$, where the $\rho_n, \sigma_n$ are fixed
functions of $\nu$ determined by the same difference equation
(\ref{DiffEqn}) with the `initial' conditions: $\rho_0(\nu) = 1, \
\rho_1(\nu) = 0$ and $\sigma_0(\nu) = 0, \ \sigma_1(\nu) = 1$ and $S_0,
S_1$ are arbitrary functions of $\nu \in (0, 4\mu_0)$. (The linearity of
the equation means that only the ratio $ \lambda(\nu) := S_1(\nu)/S_0(\nu)$ 
(say) parameterizes the general solution.)

It is clear that the arbitrary functions allow us to control the
variation of $\tilde{s}_{\mu}$ within an interval of width $4 \mu_0$. At
the integral values of $\mu/(4 \mu_0)$ corresponding to $\nu = 0$, there
is a consistency condition coming from vanishing of the highest (lowest)
order coefficient which fixes the ratio of $S_0(0)$ and $S_1(0)$. The
values of $\tilde{s}_{\mu = 4 \mu_0 n}$ are fixed (up to overall
scaling). The slowly varying class of functions will be assumed to
approximate these exact values. The continuum approximation developed
below may not be a good approximation at a {\em finite} subset of these
values corresponding to smaller $n$. 

With these remarks, we now proceed to derive consequences from the
assumption of (every where) slowly varying, approximate solutions of the
difference equation (\ref{DifferenceEqn}).

Defining $A(p) := (\Case{1}{6} \gamma \lP^2)^{\Case{3}{2}} A_{\mu}$
and substituting $\tilde{s}_{\mu}$ in terms of slowly varying $\psi(p)$ in the
difference equation (\ref{DifferenceEqn}), leads to the differential
equation,
\begin{eqnarray}
0 & = & B_0(p, p_0) \psi(p) + 4 p_0 B_-(p, p_0) \psi'(p) + 8 p_0^2 B_+(p, p_0)
\psi''(p) \mbox{\hspace{0.5cm} where,}  \label{MasterDiffEqn} \\
B_{\pm}(p, p_0) & := & A(p + 4 p_0) \pm  A(p - 4 p_0) \ ,
\mbox{\hspace{5.6cm} and} \nonumber \\
B_0(p, p_0) & := & A(p + 4 p_0) - \left(2 + 144\frac{p_0^2}{\lP^4}
\eta\right) A(p) + 
A(p - 4 p_0)  +  \left( 288 \kappa \frac{{p_0}^3}{\lP^4}\right) 
H_{m}(\mu) \nonumber 
\end{eqnarray}

In the above equation, terms involving higher derivatives of $\psi(p)$
have been neglected as being sub-leading in the context of slowly
varying solutions. This is not quite the continuum approximation referred 
to earlier since there is $\gamma$ dependence hidden inside $p_0$
appearing explicitly in the coefficients of the differential equation. 
This is also not quite the Wheeler--DeWitt equation since this equation is 
valid over the entire real line (since $p$ can take negative values 
corresponding to oppositely oriented triad) while the Wheeler--DeWitt equation
using the scale factor as independent variable is defined only for half real
line.

From the definitions of the coefficients $A, B_{\pm, 0}$, it is obvious
that under $p \to -p$ (change of orientation of the triad), $A, B_+$ and
the gravitational part of $B_0$ are all {\em odd} while $B_-$ is {\em
even}. For notational convenience we restrict to $p \ge 0$ while writing
the limiting expressions, the expressions for negative $p$ can be obtained
from the odd/even properties noted above. There are two obvious regimes to
explore which are conveniently demarcated by the scale $p_0$, namely, 
$p \gg p_0$ and $0 \le p \ll p_0$. The corresponding limiting forms for the 
coefficients $B_0, B_{\pm}$ are easily obtained. One gets,
\begin{eqnarray}
p \gg p_0 
&:&
A(p, p_0) ~ \approx ~ 3 p_0 \sqrt{p}  - \frac{1}{8}
p_0^3 \ p^{-\frac{3}{2}} \nonumber \\
&:& 
B_+(p, p_0) ~ \approx ~ 6 p_0 \sqrt{p} - \frac{49}{4}
p_0^3 \ p^{-\frac{3}{2}} \nonumber \\
&:&
B_-(p, p_0) ~ \approx ~ 12 p_0^2 \ p^{- \frac{1}{2}} 
+ o( p_0^4 \ p^{-\frac{5}{2}} ) \nonumber \\
&:& 
B_0(p, p_0) ~ \approx ~ -12 p_0^3 p^{- \frac{3}{2}} - 432 \frac{p_0^3 \sqrt{p}}{\lP^4}\eta
+ 288 \kappa \frac{{p_0}^3}{\lP^4}H_{m}(\mu) \label{LargeVolLim}\\
p \ll p_0 
&:&
A(p, p_0) ~ \approx ~ 3 p p_0^{\frac{1}{2}}  
- \frac{1}{8} p^3 p_0^{- \frac{3}{2}} \nonumber \\
&:& 
B_+(p, p_0) ~ \approx ~ 3 (5^{\frac{1}{2}} - 3^{\frac{1}{2}}) 
p p_0^{\frac{1}{2}} 
\nonumber \\
&:&
B_-(p, p_0) ~ \approx ~ 2 p_0^{\frac{3}{2}} (5^{\frac{3}{2}} -
3^{\frac{3}{2}}) \nonumber \\
&:& 
B_0(p, p_0) ~ \approx ~ (3 p p_0^{\frac{1}{2}})
( 5^{\frac{1}{2}} - 3^{\frac{1}{2}} - 2 ) - 432 \frac{p_0^{\frac{5}{2}}\ 
p}{\lP^4}\eta + 
288 \kappa \frac{{p_0}^3}{\lP^4}H_{m}(\mu) \label{SmallVolLim}
\end{eqnarray}

%
Notice that for large volume the explicit $p_0$ dependence cancels out. 
This equation corresponds to the usual continuum approximation which has no 
dependence on $\gamma$ and matches with the Wheeler--DeWitt equation
in a particular factor ordering. For small volume, the $p_0$ dependence
survives, is non-trivial and the coefficient of the first derivative terms is 
non-zero. In view of the even/odd properties of the coefficients, it follows
that the first derivative of $\psi(p)$ must vanish at $p = 0$. Further
more, even for the flat model without matter, the $B_0$ coefficient
is non-zero. Had we extrapolated the Wheeler--DeWitt equation from the
large volume form, we would not have gotten these terms. Thus the
quantum differential equation (\ref{MasterDiffEqn}) agrees with the 
Wheeler--DeWitt for large volume but differs significantly for small volume.


The small volume form of the equation in fact shows that there are two 
possible behaviors namely $\psi(p) \sim $ constant or $\psi(p)$ diverges 
as an inverse power of $p$. Neither of the indicial roots depend on the 
matter Hamiltonian. The latter solution is not slowly varying and the former
one implies that the wave function has a non-zero value at $p = 0$ and the
wave function can obviously be continued to negative $p$. Thus the
differential equation derived for slowly varying functions is both
consistent at zero volume and mimics main features of the difference
equations namely passing through zero volume and matching with 
Wheeler--DeWitt for large volume. 

In the earlier quantization of isotropic models \cite{IsoCosmo} based on 
point holonomies taking values in $U(1)$ representations (separable Hilbert
space), the decoupling of $s_0$ coefficient also implied a consistency 
condition which helped select a unique solution \cite{DynIn} from solutions 
of the Wheeler--DeWitt equation valid at large volume. With the non-separable
Hilbert space, such a condition can only result from $S^0_n$  family of
coefficients. Nevertheless, one has gotten a {\em unique} solution (up to 
normalization) thanks to the slowly varying nature of the solutions. It is
crucial here that for small $p \to 0$, the $B_-$ coefficient has a 
non-vanishing limit which forces the first derivative to vanish at $p =
0$. (If the single derivative term has been dropped, both solutions would
have been slowly varying near $p = 0$.)

In summary, with the restriction to slowly varying solutions, we have a
continuum approximation (differential equation) valid for all values of
the triad. Further more the differential equation permits a unique
solution (for each matter state) passing through $p = 0$. 

For future reference we also note that the differential equation admits
a `conserved current'. Taking imaginary part of $\psi^*$ times the
differential equation leads to,
\begin{equation}
2 p_0 B_+ \tilde{J}' + B_- \tilde{J} ~ = ~ 0~~~,~~~\tilde{J} := \psi^* \psi' - \psi (\psi^*)'
\end{equation}

Defining $J(p) := f^{-1}(p) \tilde{J}(p)$ such that $J' = 0$ determines the
function $f$. Explicitly,
\begin{eqnarray}
J(p) & = & \text{constant} \ \left( e^{\int \frac{B_-}{2 p_0 B_+}
dp}\right) \ 
\left\{ \psi^* \psi' - \psi (\psi^*)' \right\} ~~~,~~~ J' = 0 \ ;
\label{ConservedCurrent} \\
& \to & \text{constant} \ \left( p \right) \ \left\{ \psi^* \psi' - \psi
(\psi^*)' \right\}
\hspace{2.6cm} (p \gg p_0) \nonumber \\
& \to & \text{constant} \ \left( p^{\frac{8 + \sqrt{15}}{3}}\right) \ 
\left\{ \psi^* \psi' - \psi (\psi^*)' \right\}
\hspace{1.5cm} (p \ll p_0) \nonumber \ .
\end{eqnarray}

We will now go ahead with a WKB type solution and infer an effective
classical Hamiltonian.

Let $\psi(p) = C(p) e^{\Case{i}{\hbar}\Phi(p)}$. Substitution of this ansatz
in (\ref{MasterDiffEqn}) leads to a complex differential equation
involving $C(p), \Phi(p)$. The real and imaginary parts lead to,
\begin{eqnarray}
0 & = & B_0(p, p_0) + 4 p_0 B_-(p, p_0)\left\{ (\ell n C)' \right\} +
8 p_0^2 B_+(p, p_0) \left\{ - \frac{\Phi'^2}{\hbar^2} + (\ell n C)'\ ^2 
 + (\ell n C)''
\right\} \label{RealPart} \\
0 & = & 4 p_0 B_-(p, p_0) \left\{\Phi'\right\} + 8 p_0^2
B_+(p, p_0) \left\{ \Phi'' + 2 \Phi' (\ell n C)' \right\} \label{ImPart}
\end{eqnarray}

The WKB {\em approximation} consists in assuming that the amplitude $C$ is
essentially constant and the double derivatives of the phase are small
compared to the single derivatives. Consider the eq.(\ref{RealPart})
under the assumption of almost constant amplitude $C(p)$. Then this
equation is a Hamilton-Jacobi partial differential equation. These are
generally, partial differential equations involving only first derivatives with
respect to time and position and they always have an associated Hamiltonian
mechanics \cite{HamJacEqn}.
\begin{equation}
B_0(p, p_0) - 8 p_0^2 B_+(p, p_0) 
\frac{\Phi'^2}{\hbar^2} ~ = ~ 0 \label{WKBApprox}
\end{equation}

Noting that the Poisson bracket between the triad variable $p$ and the 
extrinsic curvature variable $K$ is $\Case{\kappa}{3}$, we identify
$\Phi' := \Case{3}{\kappa} K$ and  arrive at the {\em effective
Hamiltonian} as,
\begin{eqnarray}
H^{\text{eff}}(p, K, \phi, p_{\phi}) & := & - \frac{1}{\kappa}\left[ 
~ \frac{B_+(p, p_0)}{4 p_0} K^2 + \eta \frac{A(p)}{2 p_0}~ \right]
+ \frac{1}{\kappa} \left[
\left(\frac{\lP^4}{288 p_0^3}\right)
\left\{B_+(p, p_0) - 2 A(p) \right\} \right]  \nonumber \\
& & \hspace{1.0cm} + 
H_{m}(p, \phi, p_{\phi})  \label{EffHamiltonian}
\end{eqnarray}

We have multiplied by certain factors so as to get the matter Hamiltonian
term appear without any pre-factors as in the classical case. The
equation (\ref{RealPart}) of course implies $H^{\text{eff}} = 0$ and we
will interpret this as the modified Hamiltonian constraint equation. The
effective Hamiltonian is also {\em odd} under $p \to - p$ modulo the
matter term.

Note that the $K^2$ and the $\eta$ dependent terms are $o(\lP^0)$. For
large volume, the terms enclosed within the braces are vanishingly small
and the effective Hamiltonian is indeed {\em classical} (the matter
Hamiltonian receiving corrections from the inverse triad operator also
goes to the classical form without any $\lP$ dependence). For smaller
volumes, the quantum modifications are present with explicit dependence
on $\lP, p_0$. The approximation used does not quite lead to a `classical 
limit' due to explicit appearance of $\lP$.

For small volumes, the equation (\ref{SmallVolLim}) shows that $B_0,
B_+, A$ all vanish linearly with $p$ while $B_-$ goes to a positive
constant. The real and the imaginary parts of the equation,
equations(\ref{RealPart}, \ref{ImPart}), then imply that
$\Phi'$ and $C'$ both vanish, which is consistent with $\psi \sim $
constant, as deduced directly from the differential equation
(\ref{MasterDiffEqn}).

To interpret the effective Hamiltonian constraint as generating the
space-time dynamics, let us use the identification $|p| = a^2$. 
One can then obtain the extrinsic curvature $K$ from the Hamilton's equation
of motion of $p$ as,
\begin{eqnarray}
\dot{p} ~:=~ \frac{d p}{d t} & = & \frac{\kappa}{3} \frac{\partial
H^{\text{eff}}}{\partial K} ~ = ~ - \frac{B_+(p, p_0)}{6 p_0} K
\hspace{1.0cm}\mbox{or,} \nonumber \\
K & = & - 12 \left( \frac{a p_0}{B_+(p, p_0)} \right) \dot{a} 
\end{eqnarray}

The large and small volume expressions for the effective Hamiltonian and
the extrinsic curvature are,
\begin{eqnarray}
H^{\text{eff}} & \to & - \frac{3}{2 \kappa} \sqrt{p} \left( K^2 +
\eta \right) + H_{m} \label{LargeVolH} \\
K & \to & - 2 \dot{a} \hspace{9.35cm} (p \gg p_0) \label{LargeVolK} \\
H^{\text{eff}} & \to & -\frac{3}{2 \kappa}\frac{p}{\sqrt{p_0}} \left[
\left( \frac{5^{\frac{1}{2}} - 3^{\frac{1}{2}}}{2}\right) K^2 + \eta +
\frac{1}{144} \left(2 - 5^{\frac{1}{2}} + 3^{\frac{1}{2}} \right)
\left(\frac{\lP^4}{p_0^2}\right) \right] + H_{m} \label{SmallVolH} \\
K & \to & - \left(\frac{4 \sqrt{p_0}}{5^{\frac{1}{2}} - 3^{\frac{1}{2}}}\right)
\frac{\dot{a}}{a} \hspace{7.3cm} (p \ll p_0) \label{SmallVolK}
\end{eqnarray}

The large volume expressions are the same as for the classical Hamiltonian 
as expected. The small volume expressions are useful in exploring the behavior
of the effective dynamics close to the classical singularity. 

It is worth expanding on the identification $|p| = a^2$. The basic
variables of LQC are first obtained for a general homogeneous Bianchi
class A models with the Maurer-Cartan forms normalized in the usual
manner. Comparing with the metric ansatz then leads to the relations 
$|p_1| = a_2 a_3$ and cyclic. The basic variables of the {\em isotropic}
models are obtained from the above Bianchi ansatz, by putting $p_I = p, \ 
a_I = a \ \forall I$ leading to the identification above. Let us denote
this scale factor as $a_{\text{Bianchi}}$. The corresponding spatial Ricci 
scalar is $^3 R(a_{\text{Bianchi}}) = \Case{3}{2 a^2_{\text{Bianchi}}}$. On
the other hand, the standard FRW metric ansatz is so chosen that the
spatial curvature is given by $^3 R(a_{\text{FRW}}) =
\Case{6}{a^2_{\text{FRW}}}$. These two normalizations match provided
$a_{\text{Bianchi}} = \Case{a_{\text{FRW}}}{2}$. To avoid writing the
suffixes, we just note that while comparing the large volume Hamiltonian 
constraint with the standard Friedmann equation, one should use the 
replacement $a \to \Case{a}{2}$. This of course is relevant only for the
close model.

\section{Qualitative Features of the Effective Dynamics}

In the previous section we derived the effective Hamiltonian constraint
(\ref{EffHamiltonian}), using a continuum approximation keeping terms up
to {\em second} derivatives and using the WKB approximation. If we
include higher derivative terms that these would give {\em perturbative}
corrections in the large volume. Our focus is however on the small
volume regime and leading corrections which for the matter sector
include non-perturbative corrections. For our purposes the truncation to 
second derivatives suffices. 

For large volume, we already see that the effective Hamiltonian reduces to
the classical one to within terms of the order of $p^{- \Case{3}{2}}$.
We are interested in checking if the effective dynamics is non-singular
and precisely in what sense. For this it is sufficient to focus on the small
volume expressions. We will compare the classical Hamiltonian (\ref{LargeVolH})
extrapolated to small volume and the effective Hamiltonian (\ref{SmallVolH}).

Consider first the classical case. As the scale factor goes to zero, the
matter density diverges either as $a^{-3}$ for pressure-less matter or
as $a^{-4}$ for radiation. Correspondingly, $H_{m}$ either goes to
a non-negative constant or diverges as $a^{-1}$. The Hamiltonian constraint
then implies that $K$ necessarily diverges. As is well known, in
both cases the scale factor vanishes at a {\em finite} value of
synchronous time and this of course is the big bang singularity. This
also suggests a necessary condition for singular evolution: $p = 0$ should 
be reachable in finite time. Equivalently, if $p = 0$ is {\em not}
reachable in finite time, the evolution is non-singular.

Momentarily, let us {\em assume} that for some reason, the matter
Hamiltonian {\em vanishes} as the scale factor goes to zero, then $K$
must remain finite and $p = 0$ is indeed on the constraint surface.
Further more $\dot{p}$ evaluated at $p = 0$ is also zero implying that
$p = 0$ is a {\em fixed point} (rather a fixed `sub-manifold' of the
phase space of gravity and matter). The $p = 0$ trajectories of the
dynamics are then not accessible in finite time and the evolution is
non-singular. Clearly (non-) divergence of matter Hamiltonian dictates
(non-) singular evolution.

For generic (non-singular) trajectories, there are two possibilities now. 
Either (a) $p = 0$ is approached asymptotically as $t \to - \infty$ or (b)
the trajectory exhibits a bounce, $K = 0$ at a finite, non-zero $p$. For
example, in the case of scalar matter, with LQC modifications included, the 
former is realized for flat models ($\eta = 0$) while the latter is realized 
for close models ($\eta = 1$) \cite{BounceClosed}.

Consider now the quantum case. The matter Hamiltonian is guaranteed to
vanish due to the inverse volume operator definition. The arguments for
$p = 0$ being fixed point apply. However, due to the presence of $\lP^4$ 
term in eq. (\ref{SmallVolH}), there is a bounce {\em independent} of 
$\eta$ \cite{GenericBounce}. The $p = 0$ is completely decoupled from 
{\em all} other trajectories. This is exactly the same feature exhibited
by the fundamental difference equation. The exact solution $\tilde{s}_{\mu} 
= s_0 \delta_{\mu,0}$ completely decouples from all other solutions. The
bounce is then a completely generic feature of isotropic LQC. But a
bounce also provides {\em a minimum volume} for the isotropic universe
whose value is dependent on matter Hamiltonian. Such a natural, generic
scale has implications for phenomenology as well \cite{GenericBounce}.

Notice that while interpreting the effective Hamiltonian as a (modified)
constraint equation, we are keeping the kinematics of space-time (a 
pseudo-Riemannian manifold) intact. The modifications imply modification
of the dynamical aspects or equivalently of Einstein equations. To see
the modifications conveniently, let us write the Hamilton's equations in
a form similar to the usual Raychoudhuri and Friedmann equations in
terms of the FRW scale factor. 

Write the effective Hamiltonian (\ref{EffHamiltonian}) in the form,
\begin{eqnarray}
H^{\text{eff}} & = & - \frac{1}{\kappa} ( \alpha K^2 + \eta \beta ) +
\frac{1}{\kappa} \nu + H_{m} \hspace{2.0cm} {\text{where,}} \\
& & \alpha := \frac{B_+}{4 p_0} ~~,~~
\beta := \frac{A}{2 p_0} ~~,~~
\nu := \left( \frac{\lP^4}{288 p_0^3} \right) ( B_+ - 2 A )
\label{EffHam}
\end{eqnarray}
Putting $p := \Case{a^2}{4}$ ( $a$ is the FRW scale factor and $\kappa =
16 \pi G$ ) and denoting $\Case{d}{d a}$ by $'$ leads to,
\begin{eqnarray}
3 \frac{\dot{a}^2}{a^2} + 3 \frac{\eta}{a^2} & = &
\frac{16}{3} \frac{\alpha \nu}{a^4} + 3 \frac{\eta}{a^2} \left(1 -
\frac{16}{9} \frac{\alpha \beta}{a^2} \right) + \frac{16 \kappa}{3}
\frac{\alpha H_{m}}{a^4} \label{FEqn}\\
3 \frac{\ddot{a}}{a} & = & \frac{8 \alpha}{3 a^4} \left[ - 
\left\{ \left( 2 - \frac{a \alpha'}{\alpha} \right)\nu - a \nu')\right\} +
\eta
\left\{ \left( 2 - \frac{a \alpha'}{\alpha} \right)\beta - a \beta' )\right\}
\right. \nonumber \\
& & \left. \hspace{.9cm} 
- \kappa\left\{ \left( 2 - \frac{a \alpha'}{\alpha} \right)H_{m} -
a H_{m}' )\right\} \right] \label{REqn}
\end{eqnarray}
Comparing with the usual FRW equations, we {\em identify} effective
perfect fluid density and pressure as,
\begin{eqnarray}
\rho^{\text{eff}} & := & \frac{32}{3} \frac{\alpha H_m}{a^4} 
~ + ~ \frac{32}{3 \kappa} \frac{\alpha \nu}{a^4} ~ + ~ 
\frac{6}{\kappa} \frac{\eta}{a^2}
\left(1 - \frac{16}{9} \frac{\alpha \beta}{a^2} \right) \label{Density}
\\
P^{\text{eff}} & := & 
+ \frac{32}{9}\frac{\alpha}{a^4}\left\{ \left( 1 - \frac{a
\alpha'}{\alpha}\right)  H_{m} - a H_{m}' \right\} 
+ \frac{32}{9}\frac{\alpha}{\kappa a^4}
\left\{ \left( 1 - \frac{a \alpha'}{\alpha}\right)  \nu - a \nu' \right\}
\nonumber \\
& & - \frac{32}{9}\frac{\alpha}{\kappa a^4}
\eta \left\{ \left( 1 - \frac{a \alpha'}{\alpha}\right) \beta - a
\beta' + \frac{9}{16} \frac{a^2}{\alpha} \right\} 
\label{Pressure}
\end{eqnarray}
The large and small volume expressions for the effective density and
pressure are,
\begin{eqnarray}
\text{for}~ p \gg p_0 & : & \alpha \to \frac{3}{4} a ~,~ 
\beta \to \frac{3}{4} a ~, ~ \nu \to - \frac{1}{3} \lP^4 a^{-3} ; \nonumber \\
\rho^{\text{eff}}  & \to  & + 8\ a^{-3} H_m - 
\frac{8}{3 \kappa} \lP^4 a^{-6} \ , \label{LargeRho}\\
P^{\text{eff}} & \to & - \frac{8}{3} a^{-3}(a H_{m}') - 
\frac{8}{3 \kappa} \lP^4 a^{-6} \ ; \label{LargeP}\\
\text{for}~ p \ll p_0 & : & 
\alpha \to \frac{3}{16} b p_0^{-1/2} a^2 ~,~ 
\beta \to \frac{3}{8} p_0^{- 1/2} a^2 ~,~ 
\nu \to \frac{b - 2}{384} \lP^4 p_0^{-5/2} a^2 ~, \text{where}
\nonumber \\
& & b ~:=~ \sqrt{5} - \sqrt{3} \ ; \nonumber \\
\rho^{\text{eff}} & \to  & 
2 \frac{b}{\sqrt{p_0}} a^{-2} H_m - \frac{b (2 - b)}{192 \ \kappa} \lP^4
p_0^{-3} ~ + ~ \eta \frac{6}{\kappa} a^{-2} \left( 1 - 
\frac{b}{8} \frac{a^2}{p_0} \right) \ , \label{SmallRho} \\
P^{\text{eff}} & \to & 
- \frac{2}{3}\frac{b}{\sqrt{p_0}} a^{-2}\ (H_m + a H_m') +
\frac{b (2 - b)}{192 \ \kappa} \lP^4 p_0^{-3} + 
\eta \frac{2}{\kappa} a^{-2} \left(1 - \frac{3 b}{8} \frac{a^2}{p_0}
\right) \ .\label{SmallP}
\end{eqnarray}
The effective density and pressure receive contributions from the matter
sector and also the spatial curvature ($\eta$-dependent terms). Apart from 
these, the homogeneous and isotropic quantum fluctuations of the geometry 
also contribute an effective density and pressure ($\lP^4$ terms). While tiny,
these are non-zero even for large volume. Notice also that while for
large volume the spatial curvature does not contribute to the density
and pressure, for small volume it does. As a by product, we have also
obtained the density and pressure for matter, {\em directly in terms of the
matter Hamiltonian.} This is useful because currently LQC modifications
to the matter sector are incorporated at the level of the Hamiltonian
and not at the level of an action. Consequently, usual prescription for
construction of the stress tensor and reading off the density and
pressure is not available. These definitions of course automatically satisfy 
the conservation equation: $a \rho' = -3 (P + \rho)$.

Let us consider the vacuum sector, $H_{m} = 0$. The more general case of 
presence of scalar field matter is discussed in \cite{GenericInflation}. Even for
the flat model ($\eta = 0$), the $o(\lP^4)$ terms contributes to the
effective density and pressure. This term in the effective Hamiltonian,
$W_{qg} := \Case{\nu(p, p_0)}{\kappa}$, is a `potential' term and will be
referred to as the {\em quantum geometry potential}. It is actually of order 
$\sqrt{\Case{\hbar}{\kappa}}$ after expressing $p, p_0$ in terms of the 
$\mu, \mu_0$.
\begin{figure}[htb]
\begin{center} 
\includegraphics[width=12cm]{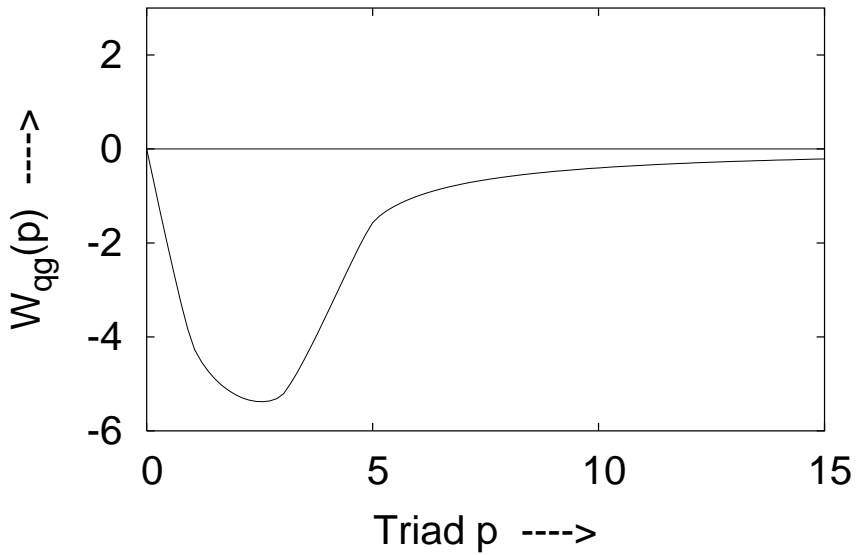}
\end{center}
\caption{ The quantum geometry potential. The triad variable $p$ is in
units of $p_0$ while the potential $W_{qg}(p)$ is in units of
$\Case{1}{288 \kappa}\lP^4 p_0^{- 3/2}$. }
\label{QPotlFig}
\end{figure}
It is easy to see that the quantum geometry potential is {\em odd} under $p \to
-p$ (since $B_+$ and $A$ are odd) and for $p > 0$ it is {\em negative-definite}.
Its plot is shown in the figure \ref{QPotlFig}. This immediately implies that
in the absence of matter (and cosmological constant), all the three terms in
the effective Hamiltonian must be individually zero which is not possible for
the quantum geometry potential. In other words, there is no solution 
space-time. This is in contrast to the purely classical Hamiltonian which 
does give the Minkowski space-time \footnote{The Minkowski space-time here
refers to Riemann flat metric regardless of its global topology. In the 
cosmological context, the spatial slice is always compact.} as a solution 
for the flat case ($\eta = 0$). This is also apparent from the non-zero 
value of the effective density which prevents the Minkowski solution.

This feature can also be understood in the following manner. The
differential equation has a unique solution (which is slowly varying
every where). This solution is however purely real and does not admit a WKB
form. Consequently, the universe does not admit any `classically allowed
region' in the WKB sense and thus also does not exhibit a classical
behavior. Once matter is included, we have again a unique, real
solution of the differential equation {\em for every matter state}. We
can now have complex linear combinations admitting possibility of
regimes of WKB form and corresponding classical behavior.

It is straight forward to write down a Lagrangian from the $H^{\text{eff}}$ as, 
\begin{eqnarray} 
L^{\text{eff}}(p, \dot{p}) & := & \frac{3}{\kappa} K \dot{p} - 
H^{\text{eff}}(p, K) \nonumber \\
& = & - \frac{1}{\kappa}\left[ 9 \frac{p_0}{B_+} \dot{p}^2  
 - \ \eta \frac{A(p)}{2 p_0}~ 
+ \left(\frac{\lP^4}{288 p_0^3}\right)
\left\{B_+(p, p_0) - 2 A(p) \right\}
\right] \label{EffLagrangian}
\end{eqnarray}

It would be interesting to see if this Lagrangian can be obtained from a 
specialization of a generally covariant action to homogeneous, isotropic 
metric. 

\section{Conclusions and Outlook}

The results of this paper are based on two essential ingredients: the
proposal of a continuum approximation for all volumes exploiting the 
non-separable nature of the kinematical Hilbert space and
the derivation of the effective Hamiltonian via the WKB route. 

The continuum approximation step leads to a differential equation for a
(still) quantum wave function, $\psi(p)$. This equation matches with the
Wheeler--DeWitt equation for large volume and has important deviations
(the first derivative terms) from it at small volumes. These deviations 
allow continuation of the wave function through zero volume just as the 
fundamental difference equation does. For slowly varying solutions, it picks
out the `boundary' condition $\Case{d \psi}{d p}(0) = 0$. Again this is
analogous to unique solution (per matter state) picked out by the
difference equation obtained from the $U(1)$ point holonomies in the
earlier work \cite{DynIn}. Thus the essential features of the fundamental
difference equation namely non-singular quantum evolution with semi-classical
limit are captured by the continuum differential equation.

The effective dynamics specified by the effective Hamiltonian deduced via
the WKB approximation also reflects these features. The effective
dynamics is non-singular, captures the decoupling of the $\tilde{s}_{\mu} 
= s_0 \delta_{\mu, 0}$ exact solution of the difference equation by making
the classical $p = 0$ trajectory decouple and reduces to the usual
classical dynamics of general relativity for large volumes. Since the 
essential features of quantum dynamics are now captured in classical 
geometrical terms, the effective dynamics is more reliable than the usual 
one and one can now simply work with the effective dynamics to do 
phenomenology. Already, at the qualitative level, one sees that all non-trivial
evolutions necessarily show a bounce providing a natural scale for say, density
perturbations and their power spectra.

Since the approach draws on the WKB method, a few remarks on the
interpretational aspects are in order.

From the continuum quantum dynamics, with the WKB approximation, one
obtains a Hamilton-Jacobi equation. As a mathematical result, any 
Hamilton-Jacobi differential equation has an associated Hamiltonian 
mechanics \cite{HamJacEqn} and corresponding trajectories. In our
context, we are interpreting these trajectories as possible evolutions 
of the isotropic universe. There is at least an implicit implication (or
assumption) that a quantum system executing a WKB approximable quantum 
motion {\em physically} exhibits a classical motion governed by the 
Hamiltonian associated with the corresponding Hamilton-Jacobi equation.
The justification for this comes from noting that for large volume we
expect the universe to exhibit classical behavior and there it is
indeed governed by the associated Hamiltonian. For how small volumes can
we assume this expectation? This question is naturally related to the domain
of validity of the WKB approximation.

One expects the WKB approximation (slow variation of the phase and
almost constancy of the amplitude of the wave function) to break down
closer to the classically indicated singularity at zero volume. Noting
that the differential equation is local in $p$ and its solutions are
also local solutions (valid in open intervals in $p$), we can begin with
a WKB approximable solution valid in the larger volume regime and
attempt to extrapolate it to smaller and smaller volumes. All through
these extensions, one will have the effective Hamiltonian with its
associated trajectories which can access the values of $p$ in these
intervals. The effective Hamiltonian constraint defines a submanifold of
the phase space and all trajectories must lie on this. The range of
configuration space variables (eg $p$ in our case) allowed by the
submanifold defines `classically allowed region'. As is well known from
the usual examples in non-relativistic quantum mechanics, the WKB 
approximation breaks down at the `turning points'. These correspond to the 
boundary of classically allowed region which therefore demarcates the 
domain of validity of WKB approximation. Clearly, when such a boundary
is reached by a trajectory, it must turn back. This is of course the
bounce ($\dot{a} = 0$). The expectation that WKB breaks down at non-zero
volume translates into the expectation of a bounce occurring at non-zero scale
factor. The bounce can thus be understood as the smallest volume (or scale
factor) down to which one may use the classical framework with some 
justification but below it one must use the quantum framework. More details
of the bounce picture as well as its genericness are discussed in 
\cite{GenericBounce}.

Further justification comes from other known examples. For example, solutions
of the Maxwell equations in the eikonal approximation can be understood in
terms of the normals to the wave fronts which follow null geodesic. The 
interpretation that this actually reflects rectilinear motion of light, may be
justified by noting that the Poynting vector (energy flow) is also in the 
same direction as the normals. Likewise, in the context of usual Schrodinger
equation of particle mechanics, the conserved probability current also points
along the normals to the wave fronts giving credence to the interpretation
that a quantum state of the WKB form realizes motion of a particle (or
wave packet) governed by the associated Hamiltonian mechanics. 
In both these examples, further inputs other than the mathematical association 
between Hamilton-Jacobi differential equation and a Hamiltonian system, 
seem to be  needed to understand the physical realization of the Hamiltonian
system.

Interestingly, the equation (\ref{MasterDiffEqn}) does admit a conserved 
current (\ref{ConservedCurrent}) which indeed is proportional to the
gradient of the WKB phase. Whether this could be used for guessing
physical inner product vis a vis a probability interpretation remains to
be seen.
\begin{acknowledgments}
We thank Martin Bojowald for helpful, critical comments.
\end{acknowledgments}


\end{document}